\documentclass{article}

    \PassOptionsToPackage{numbers, compress}{natbib}

 \usepackage[dblblindworkshop, final]{neurips_2025}

\usepackage[utf8]{inputenc} 
\usepackage[T1]{fontenc}    
\usepackage{hyperref}       
\usepackage{url}            
\usepackage{booktabs}       
\usepackage{amsfonts}       
\usepackage{nicefrac}       
\usepackage{microtype}      
\usepackage{xcolor}         
\usepackage{graphicx}
\usepackage{tcolorbox}
\usepackage{tabularray}
\usepackage{caption}
\usepackage{booktabs} 
\usepackage{tabularray}
\usepackage{nicematrix} 
\usepackage{adjustbox} 
\usepackage{wrapfig}
\usepackage{subcaption} 
\usepackage{float} 

\title{EvoMem: Improving Multi-Agent Planning with Dual-Evolving Memory}

\author{%
  Wenzhe Fan\textsuperscript{ 1,$\dagger$} \hspace{2em}
  Ning Yan\textsuperscript{ 2} \hspace{2em}
  Masood Mortazavi\textsuperscript{ 2} \\
\\
  \textsuperscript{1} University of Illinois Chicago \\ 
    \textsuperscript{2} Futurewei Technologies  
}

\workshoptitle{Bridging Language, Agent, and World Models for Reasoning and Planning}

\begin{document}

\maketitle

\begingroup
\renewcommand{\thefootnote}{\fnsymbol{footnote}}
\footnotetext[2]{Work done during the author’s internship at Futurewei Technologies. Correspondence to: \texttt{wfan23@uic.edu}}
\endgroup


\begin{abstract}
Planning has been a cornerstone of artificial intelligence for solving complex problems, and recent progress in LLM-based multi-agent frameworks have begun to extend this capability.
However, the role of human-like memory within these frameworks remains largely unexplored.
Understanding how agents coordinate through memory is critical for natural language planning, where iterative reasoning, constraint tracking, and error correction drive the success.
Inspired by working memory model in cognitive psychology,
we present EvoMem, a multi-agent framework built on a dual-evolving memory mechanism. 
The framework consists of three agents (Constraint Extractor, Verifier, and Actor) and two memory modules: 
Constraint Memory (CMem), which evolves across queries by storing task-specific rules and constraints while remains fixed within a query, 
and Query-feedback Memory (QMem), which evolves within a query by accumulating feedback across iterations for solution refinement.
Both memory modules are reset at the end of each query session.
Evaluations on trip planning, meeting planning, and calendar scheduling show consistent performance improvements,
highlighting the effectiveness of EvoMem.
This success underscores the importance of memory in enhancing multi-agent planning.
\end{abstract}

\vspace{-0.5em}
\section{Introduction}
\vspace{-0.5em}
Planning is a fundamental cognitive ability that involves generating sequences of actions and reasoning about future states \citep{jiao2024learning, wang2024planning}.
However, prior work has shown that large language models (LLMs) often struggle with planning tasks, especially those requiring multi-step reasoning or long-range dependencies \citep{kambhampati2024llms, valmeekam2023planning, shojaee2025illusion}.
At the same time, the crucial role of memory in supporting computational reasoning, planning, and other cognitive functions has been widely recognized across computing, computational neuroscience, and psychology \citep{schuurmans2023memory, momennejad2024memory, armeni2024transformer, gollwitzer2025psychology, brown2020stress}. 
Yet, while memory has been understood as useful to reasoning and adaptation, its structure and mechanisms in the natural language planning tasks remain largely underexplored in LLM-based multi-agent frameworks.

Inspired by working memory model in cognitive psychology \citep{baddeley2012working},
we propose EvoMem, a multi-agent framework that explicitly incorporates the dual-evolving memory into the planning process. 
The framework assembles three agents (Constraint Extractor, Verifier, and Actor) and two memory modules.  
Given a query, the Constraint Extractor identifies task-specific rules and constraints, establishing the Constraint Memory (CMem).
This fixed memory anchors the Actor’s solution generation across iterations and updates only when a new query is introduced.
The Verifier evaluates the Actor’s output and provides feedback for subsequent iterations, forming the Query-feedback Memory (QMem).
This dynamic memory evolves within a single query, enabling solution refinement throughout the ongoing task.
CMem and QMem are reset after each query.

EvoMem is conceptually aligned with Baddeley’s multi-component theory of working memory first proposed in 1974 \citep{baddeley2012working} (Details shown in App. \ref{app:wm} ). 
CMem plays a role analogous to the phonological loop by maintaining all constraints in a stable, verbalized form across rounds. 
QMem functions similarly to the sketchpad, accumulating intermediate outputs, verifier feedback, and error signals into a coherent multi-round representation. 
The iterative Actor–Verifier process serves as a central executive that selectively attends to CMem, updates QMem, and directs the next reasoning step.
This cognitive framework explains why combining long-lived constraints with dynamically updated, per-round information leads to a more consistent and effective planning process.

We evaluate EvoMem on the NaturalPlan benchmark, covering trip planning, calendar scheduling, and meeting planning tasks. 
Using Gemini-1.5-Pro \citep{team2024gemini} as the primary backbone, EvoMem outperforms strong baselines, achieving average gains of +11.17\% in trip planning, +2.56\% in calendar scheduling, and +3.76\% in meeting planning. To assess model-agnostic performance, we further test on DeepSeek V3 \citep{liu2024deepseek} and GPT-4.1-mini \citep{achiam2023gpt}, and observe consistent improvements across models.

In summary, our contributions are as follows:
First, we present EvoMem, a multi-agent framework that integrates specialized agent roles with dual-evolving memory mechanism to address complex natural language planning tasks. 
Second, we demonstrate that EvoMem achieves state-of-the-art performance on the NaturalPlan benchmark, validating the effectiveness of its memory-centric design. 
Finally, we show that for complex planning tasks, maintaining per-query memory is sufficient to capture iterative reasoning and substantially improve performance.


\vspace{-0.5em}
\section{EvoMem}
\begin{figure}[!t]
    \centering
    \includegraphics[width=0.8\textwidth]{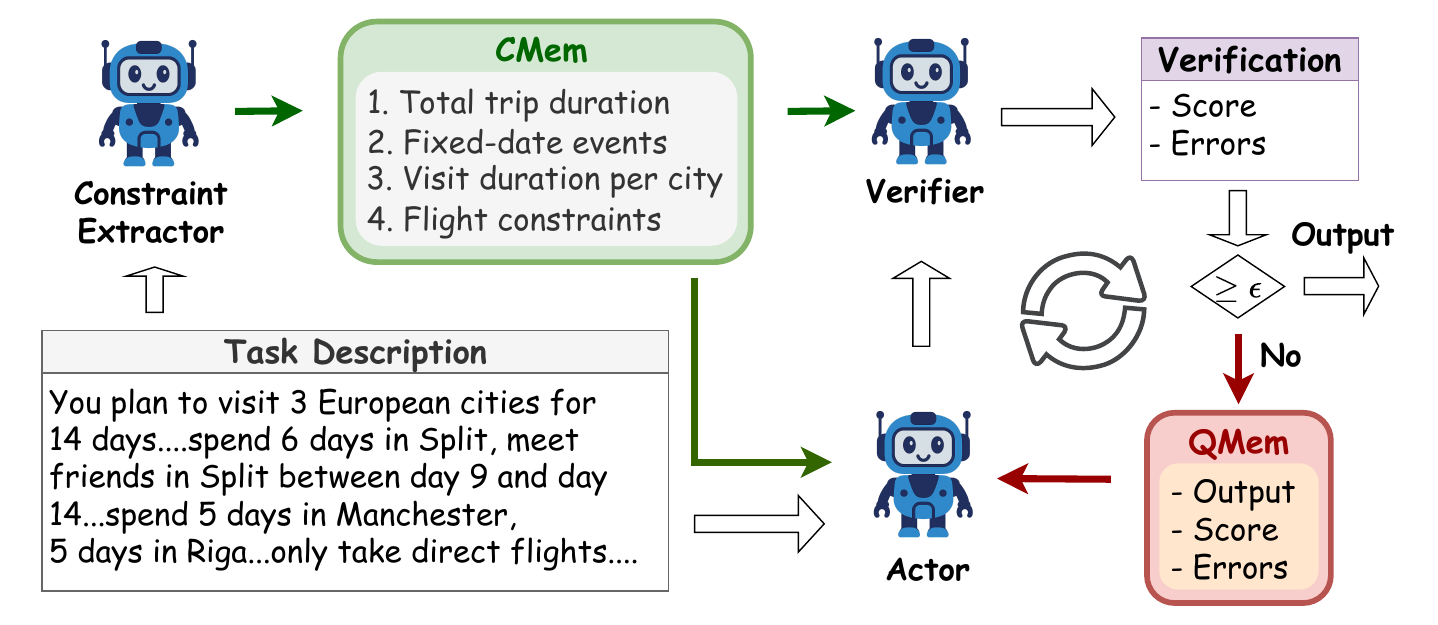}
    \caption{The dual-evolving memory mechanism in EvoMem: CMem evolves across queries to establish fixed, task-specific constraints for the actor and verifier. QMem evolves within a single query by recording failed attempts (solution, score, errors), providing iterative feedback until the solution is verified or the turn limit is reached.} 
    \label{fig:EvoMem}
\end{figure}

\vspace{-0.5em}
We propose EvoMem, a multi-agent framework for solving complex planning tasks through an iterative self-correction process. As illustrated in Figure~\ref{fig:EvoMem}, the framework employs two complementary evolving memory modules, i.e., CMem and QMem, to guide reasoning and ensure that solutions progressively align with all specified constraints.

\vspace{-0.5em}
\subsection{Multi-Agent Design}
\vspace{-0.5em}
\textbf{Constraint Extractor} \quad
Constraints are the rules and limitations that a valid plan must satisfy. 
For instance, in trip planning tasks, constraints may involve the total trip duration, fixed-date events (e.g., weddings or conferences), city-specific stays, and available flights. 
Prior work \citep{parmar2025plangen} indicates the criticality of constraint identification for reliable plan verification.
The constraint extractor derives task-specific constraints from the problem description, capturing the essential conditions that any candidate plan must satisfy. 
These constraints are then appended to the actor’s prompts, ensuring its reasoning remains grounded in the core requirements of the task.

\textbf{Verifier} \quad
The verifier evaluates each solution against the extracted constraints and facilitates refinement through QMem-mediated feedback.
This agent provides two types of output:
(i) \textit{Reward Score (0-100)} that measures plan quality based on constraint satisfaction, and (ii) \textit{Feedback} (including constraint violations) which the agent deposits in QMem for future iterations. To ensure full compliance, only plans with a perfect score of 100 are accepted.

\textbf{Actor} \quad
The actor generates a solution for a given query. 
The actor takes the query and the extracted constraints (CMem), incorporating feedback from QMem in subsequent turns if an earlier turn fails.
Prompts and examples for these three agents are given in App. \ref{app:prompts} and App. \ref{app:examples}.

\vspace{-0.5em}
\subsection{Memory Modules}
\vspace{-0.5em}
Given a query, our framework generates a solution through the iterative process of up to $T$ turns.
Planning is jointly guided by the Constraint Memory (CMem), which adapts across queries but remains fixed within a single query, and the Query-feedback Memory (QMem), which evolves across iterations within the same query. Together, these two modules form what we call ``a dual-evolving memory'' mechanism. CMem and QMem are reset at the end of each query session.

\textbf{Constraint Memory (CMem)} \quad
CMem stores the constraints extracted from the query and is appended to the actor’s prompt at every turn. 
It remains fixed throughout the multi-turn refinement process, ensuring that the actor consistently adheres to the core requirements of the problem. 
The importance of CMem is further validated in our ablation study (Sec. \ref{sec:ablation}).

\textbf{Query-feedback Memory (QMem)} \quad
QMem is a dynamic memory module that evolves across iterations by recording information from failed attempts in earlier turns.
Whenever a candidate solution violates constraints, its solution, score, and identified errors are logged in QMem. 
This memory evolves across at most $T$ iterations, guiding subsequent refinements. 

\vspace{-0.5em}
\subsection{Proposed Framework}
\vspace{-0.5em}
The planning process for a given query unfolds over $T$ iterations, with each turn progressively refining the solution under the guidance of memory.

\textbf{\textit{Constraint Extraction}}:
First, the constraint extractor identifies all relevant constraints from the query and records them in CMem. 
This module remains fixed throughout the query session, 
serving as a stable anchor for all subsequent iterations.
\textbf{\textit{Solution Generation}}:
The actor then generates an initial solution based solely on the constraints in CMem. 
In later iterations, 
solution generation is enhanced by also incorporating the accumulated feedback stored in QMem.
\textbf{\textit{Verification}}:
Next,
the verifier evaluates the candidate solution against the stored constraints in CMem, producing a reward score and listing any violated constraints. 
If no violations are found, the solution is accepted. 
\textbf{\textit{Self-Correction}}:
If the verifier finds violations, the framework enters the Self-Correction loop. 
The failed solution, its score, and the specific errors are all logged as a new entry in QMem. 
In subsequent iterations, the actor leverages both the fixed constraints in CMem and the accumulated feedback in QMem to refine its output. 
The cycle of generation, verification, and memory update continues until either a valid solution is found or the maximum number of $T$ turns is reached.
\textbf{\textit{Reset}}:
Finally, both CMem and QMem are cleared before a new query begins, 
preventing any cross-contamination.

\begin{figure}[!tp]
    \centering
    \begin{subfigure}[t]{0.48\textwidth}
        \centering
        \includegraphics[width=0.9\linewidth]{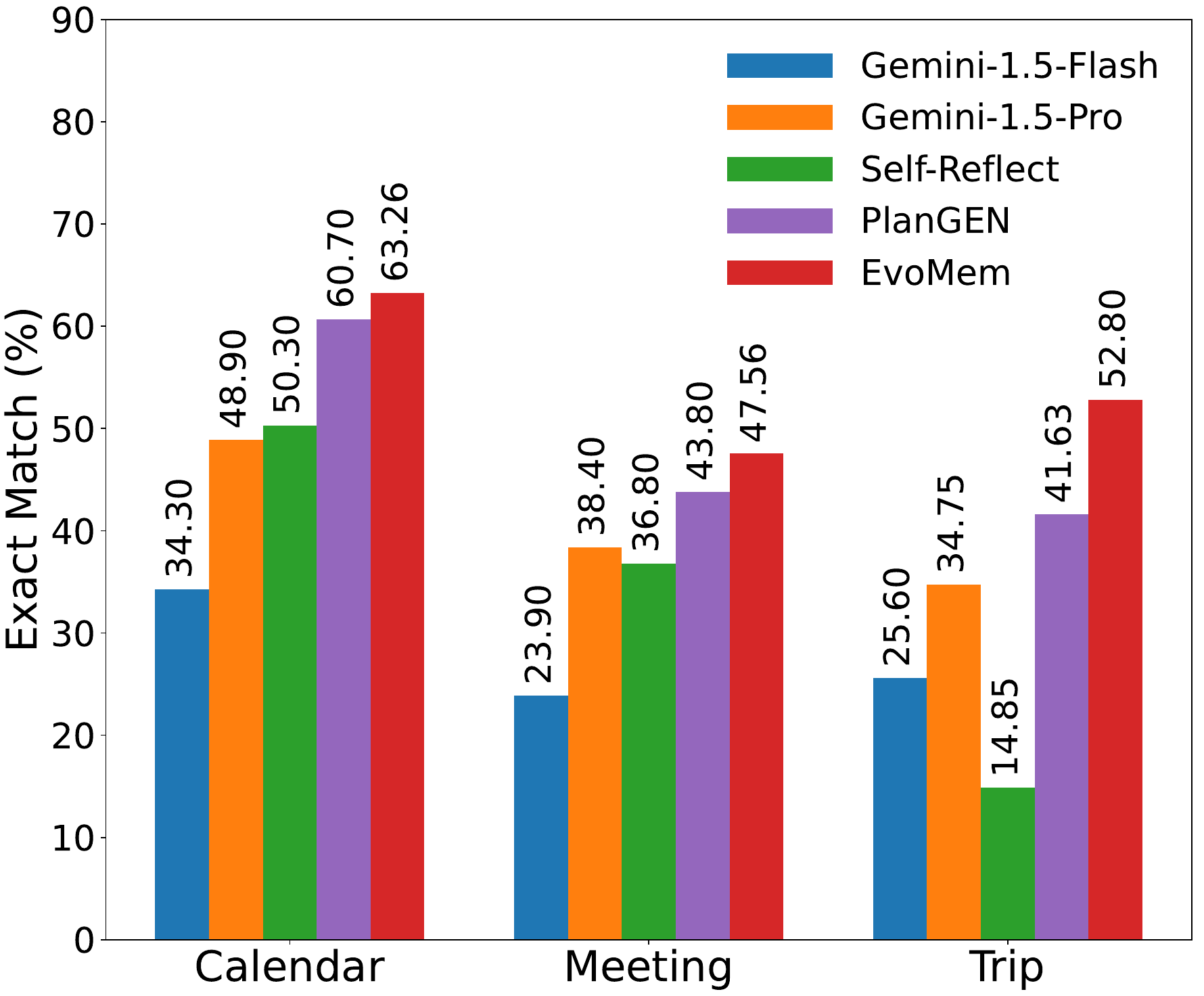}
        \caption{Performance}
        \label{fig:main_result}
    \end{subfigure}
    \begin{subfigure}[t]{0.48\textwidth}
        \centering
        \includegraphics[width=0.9\linewidth]{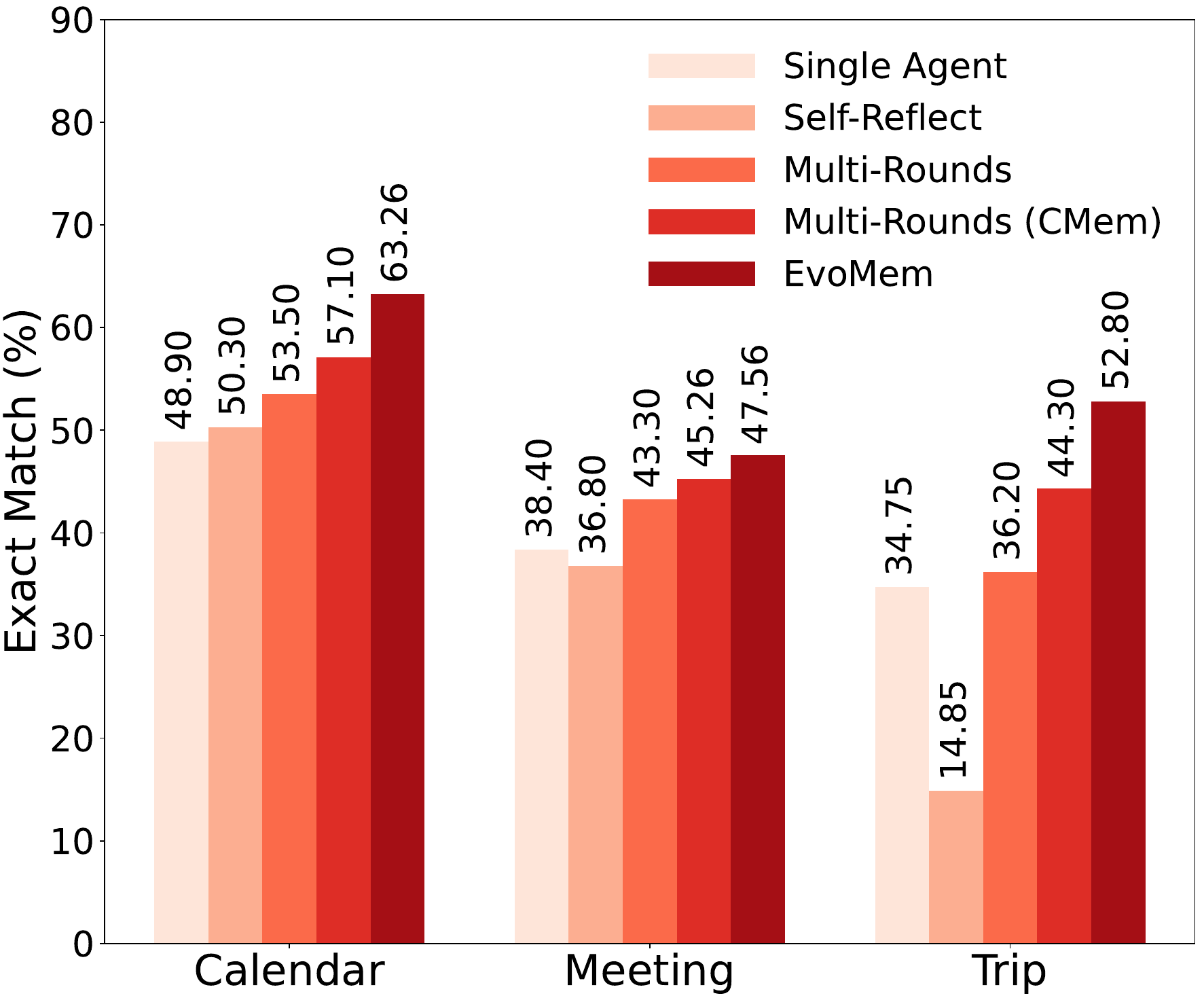}
        \caption{Effect of different components}
        \label{fig:ablation}
    \end{subfigure}
    \caption{\small{(a) Performance comparison of EvoMem against other baseline methods. (b) Performance comparison of EvoMem gradually adding one more component. (All experiments are conducted with Gemini-1.5-Pro.)}} 
    \label{fig:results}
\end{figure}

\vspace{-0.5em}
\section{Experiment}
\vspace{-0.5em}

\subsection{Main Result}
\vspace{-0.5em}
We conduct experiments on three datasets in NaturalPlan \citep{zheng2024natural}: 1,000 instances each for Calendar Scheduling and Meeting Planning, and 1,600 instances for Trip Planning. For comparison, we adopt three baselines: (i) Zero-shot CoT \citep{kojima2022large}, (ii) Self-Reflect \citep{parmar2025plangen}, where the same model iteratively refines outputs through self-reflective feedback loops, and (iii) PlanGen \citep{parmar2025plangen}, a strong method that achieves state-of-the-art results in multi-agent frameworks across these datasets. In EvoMem, we set the temperature of the constraint extractor to 0.1, the verifier to 0, and the actor to 0.7. We set the maximum number of iterations $T$ to 5.

As shown in Figure \ref{fig:main_result}, EvoMem achieves the highest exact match scores \citep{zheng2024natural} across all tasks averaged over 5 runs: 63.26$\pm$0.41\% (Calendar), 47.56$\pm$0.32\% (Meeting), and 52.08$\pm$0.12\% (Trip). 
All agents in these experiments are powered by Gemini-1.5-Pro.
These results highlight the effectiveness of EvoMem in diverse natural language planning tasks and establish a new state-of-the-art among multi-agent framework methods.
They further confirm the crucial role of query-specific memory in advancing the planning capabilities of LLM-based multi-agent systems.

\begin{table}[!htp]
\centering
\begin{adjustbox}{width=\textwidth} 

\begin{tblr}{
  colspec={*{15}{c}},
  hlines = {0.05em},
  vline{2-12} = {-}{},
  hline{1,5} = {-}{0.08em},
  row{1-Z} = {valign=m},
}
\textbf{Methods}               & \textbf{Trip} & \textbf{Calendar} & \textbf{Meeting} & \textbf{Methods}            & \textbf{Trip} & \textbf{Calendar} & \textbf{Meeting} & \textbf{Methods}            & \textbf{Trip} & \textbf{Calendar} & \textbf{Meeting} \\
Gemini-1.5-Pro                 &  34.75\%      &    48.9\%         &        38.4\%    & DeepSeek v3                 &  38.5\%       &   57\%            &      46.3\%      & GPT-4.1-mini                 &    24.75 \%   &    45.6\%         &  37.2\%        \\
{EvoMem\\(Gemini-1.5-Pro)}    &  53.5\%       &    63.26\%        &        47.56\%   & {EvoMem\\(DeepSeek v3)}    &  49\%  &   61.1\%          &   50.2\%      & {EvoMem\\(GPT-4.1-mini)}    &    39.1\%     &    66.8\%         &  33.71\%       \\
\end{tblr}
\end{adjustbox}
\vspace{6pt}
\caption{\small{Performance comparison of EvoMem with three base LLM models on three different planning tasks.}} 
\label{table:llms}
\end{table}

\subsection{Ablation Study}
\label{sec:ablation}
\vspace{-0.5em}

\textbf{Effect of Different Components} \quad
Figure \ref{fig:ablation} presents the ablation study evaluating the contribution of each component in our framework. 
We begin with a ``Single Agent'' setting, where only the actor is used. 
Adding ``Self-Reflect'' enables multi-round self-evaluation but produces mixed results, emphasizing the necessity of an explicit verifier guided by rules. 
Incorporating both the constraint extractor and verifier in the ``Multi-Rounds'' yields consistent improvements, underscoring the importance of constraint-based verification. 
Performance is further enhanced by introducing the CMem module, which ensures that the actor remains aligned with the query’s constraints across iterations. 
Finally, adding QMem, which accumulates errors from previous attempts, provides additional gains, demonstrating the benefit of evolving,  query-feedback memory in iterative planning.

\textbf{Effect of Different LLMs} \quad
The results in Table \ref{table:llms} demonstrate the impact of EvoMem across different LLM backbones. 
With Gemini-1.5-Pro, EvoMem achieves the strongest overall performance, improving Trip by +18.75\%, Calendar by +14.36\%, and Meeting by +9.16\% relative to the base model. 
Comparable gains are observed with DeepSeek v3, where EvoMem adds +10.5\% on Trip, +4.1\% on Calendar, and +3.9\% on Meeting. 
Even with the smaller GPT-4.1-mini, EvoMem substantially enhances performance, particularly on Calendar (+21.2\%), though Meeting shows a slight decline (–3.5\%). 
While absolute performance varies across different backbones, EvoMem consistently delivers relative improvements, indicating that its memory-driven mechanism generalizes well across models. 


\textbf{Effect of Self-Correction Rounds} \quad 
We observe that the performance remains stable across maximum iteration number $T \in \{3, 5, 7\}$, with an average of $53.284\% \pm 0.357 \%$.
A cost-benefit analysis on the iteration cap $K$ (with $T=7$) revealed diminishing returns, as each additional iteration yields fewer new successes.
The first three iterations ($K=3$) are highly effective, successfully resolving 93.7\% of all solvable tasks while using only 67.88\% of the total queries.
Extending the cap from $K=3$ to $K=5$ yields 45 additional successes (from 93.7\% to 97.6\%) for only ~2.8\% more dataset coverage.
Going from $K=5$ to $K=7$ nets just 28 more successes to complete the dataset, the most expensive part for the smallest gain.
These findings indicate that $T=5$ as the optimal balance point of performance and efficiency, while $T=3$ is a decent choice for cost-sensitive applications. The details of this study are shown in App.~\ref{app:self_reflect_rounds}.

\textbf{Effect of Temperature} \quad
We first fix the temperatures of verifier and constraint extractor while varying the actor’s temperature from 0.7 to 0.5 and 0.3. 
We then evaluate two additional settings: all agents with temperature 0 (fully deterministic) and all agents with temperature 1 (Gemini default). The results show a variance of only 0.256\%, indicating that the framework is highly robust to temperature settings.


\vspace{-0.5em}
\section{Conclusion and Future work}
\vspace{-0.5em}
In this work, we introduced EvoMem, a dual-evolving memory framework for natural language planning. 
EvoMem integrates three agents (Constraint Extractor, Actor, and Verifier) with two evolving memory modules: a Constraint Memory (CMem) that evolves at the query level to set fixed, task-specific constraints for a given task, 
and a Query-feedback Memory (QMem) that evolves at the iteration level within a query by accumulating multi-turn feedback. 
Experiments on trip planning, meeting planning, and calendar scheduling show consistent gains across diverse LLM backbones, demonstrating the value of EvoMemo and further emphasizing the importance of memory in advancing multi-agent planning.
The updated working memory theory incorporates long-term memory, suggesting promising directions for extending EvoMem toward richer long-term or learned workflow memories to support more complex planning and reasoning tasks.

\bibliographystyle{plainnat}
\bibliography{reference}

\newpage

\appendix
\section{Related Work}
\label{app:related_work}
\textbf{LLM Agents for Planning} \quad
Recent research has established that Large Language Models (LLMs) struggle to solve planning tasks directly from task descriptions \citep{kambhampati2024llms, shojaee2025illusion, valmeekam2023planning}.
This limitation has spurred the development of various natural language planning benchmarks and environments to study the problem comprehensively \citep{zheng2024natural, bohnet2024exploring, xie2024travelplanner, valmeekam2023planbench}. 
In this work, we evaluate our approach on NaturalPlan \citep{zheng2024natural}, which includes trip planning, calendar scheduling, and meeting planning tasks.

In response to this challenge, a significant body of work has focused on creating LLM-based agent frameworks to enhance planning capabilities. Some approaches are tailored to specific domains \citep{liu2024tool, wu2024copilot}, 
Other methods \citep{zhu2024knowagent, xie2024human} focus on integrating external information to guide the planning process. 
Another line of research concentrates on optimizing the prompts themselves \citep{chen2024reprompt, wang2023promptagent}. 

The current SOTA LLM-based agent framework on natural language planning benchmarks is held by PlanGen \citep{parmar2025plangen}. 
While recent works continue to make progress, they have primarily focused on innovations within single-agent frameworks \citep{lee2025evolving, chen2025sets}. 

Our work contributes to this field by investigating a different axis of improvement: the role of the memory mechanism. We demonstrate that by incorporating memory into a simple multi-agent framework, our approach achieves better result than PlanGen on NaturalPlan benchmark.

\textbf{Memory for LLM Agents} \quad
The importance of memory in computational, planning and other cognitive tasks has been well established, e.g., by computing, computational neuroscience, cognition and psychology researchers \citep{schuurmans2023memory, momennejad2024memory, armeni2024transformer, ClarkSuperSizing, gollwitzer2025psychology,  brown2020stress}.
Memory mechanisms have become central to advancing LLM-based agents, as effective planning and reasoning often require retaining and reusing information beyond a single context window. 
Early efforts focused on memory utilization \citep{chhikara2025mem0, wu2024copilot, mei2024aios, zhong2024memorybank, liu2023think, wang2023enhancing}, using short-term and long-term memory to support information storage and recall. 
More recent work has explored agentic memory \citep{xu2025mem, tang2024enhancing}, where agents dynamically organize memories, establish new connections, and evolve their knowledge with experience. 
Another important direction is workflow memory \citep{wang2024agent}, which captures and reuses common routines to guide future tasks and improve efficiency.
However, none of the above methods investigate the effectiveness of memory mechanisms in natural language planning tasks. 
Our framework specifically addresses this gap.

\section{Working Memory Model in Cognitive Psychology}
\label{app:wm}
\begin{figure}[h]  
    \centering
    \includegraphics[width=0.5\linewidth]{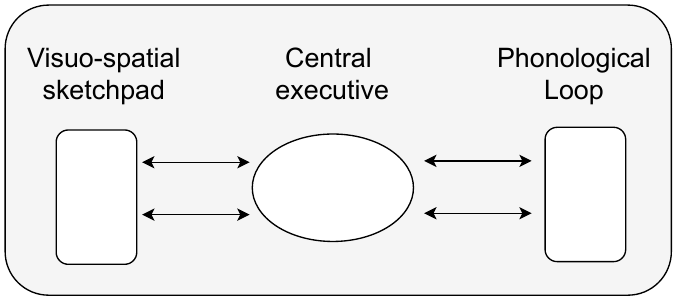}
    \caption{Working memory model proposed in 1974 \citep{baddeley2012working}.}
    \label{fig:wm_model}
\end{figure}

Figure~\ref{fig:wm_model} shows the classic working memory model with three parts: the central executive, the visuo-spatial sketchpad, and the phonological loop.

The central executive is like the ``manager" of your mind’s working memory. It controls your attention, decides what matters, and coordinates the other components.
The visuo-spatial sketchpad is the ``inner eye".
It holds visual and spatial information—like imagining a map, remembering where objects are, or picturing a diagram—and becomes active whenever you visualize something in your head.
The phonological loop is the “inner voice”.
It keeps words or sounds in your mind for a short time.
For example, you repeat a phone number to yourself so you do not forget it.
\clearpage

\section{Ablation: Effect of Self-Correct Rounds}
\label{app:self_reflect_rounds}

\begin{table}[ht!]
\centering
\begin{tblr}{
  colspec = {*{7}{c}},        
  hlines = {0.05em},
  vline{2-7} = {-}{0.05em},
  hline{1,7} = {-}{0.08em},
}
{iteration\\cap K} & {\#queries \\ finished \\  $\leq$ K} & \% of dataset & {\#Success\\(score=100)\\ $\leq$ K} & {\% of all \\ successes } & {+ Successes vs\\prev. cap} & {Success\\remaining} \\
3                  & 1,086                                       & 67.88\%       & 1,086                                           & 93.7\%              & -                           & 73 (6.3\%)           \\
4                  & 1,111                                       & 69.44\%       & 1,111                                           & 95.86\%             & +25                         & 48 (4.14\%)          \\
5                  & 1,131                                       & 70.69\%       & 1,131                                           & 97.58\%             & +20                         & 28 (2.42\%)          \\
6                  & 1,146                                       & 71.63\%       & 1,146                                           & 98.88\%             & +15                         & 13 (1.12\%)          \\
7                  & 1,600                                       & 100\%         & 1,159                                           & 100\%               & +13                         & 0 (0\%)              
\end{tblr}
\vspace{6pt}
\caption{Success coverage (score=100) vs. iteration gap $K$ (maximum iteration number $T=7$).}
\label{table:self_reflect_rounds}
\end{table}

\section{Details on LLM Agents}
\label{app:prompts}

\begin{tcolorbox}[title=Constraint extractor Prompt,
  colback=gray!0,   
  colframe=black!65, 
  fonttitle=\bfseries,
  width=1.1\textwidth,
  valign=center, 
]
You are an expert constraint extraction agent for planning problems. 
Your task is to extract only structured constraints in a clean, concise, and strictly formatted style."\\

Extract constraints using only the format below: \{Output Format\} \\

(Optional) There is information might include: \\
\{You may provide the specific constraints you want\}\\
\{You may provide example query and corresponding extracted constraints\} \\
\textbf{Query}:\{query\}
\end{tcolorbox}

\vspace{3em}

\begin{tcolorbox}[title=Verifier Prompt,
  colback=gray!0,   
  colframe=black!65, 
  fonttitle=\bfseries,
  width=1.1\textwidth,
  valign=center, 
]
You are a meticulous verifier responsible for evaluating trip plans against a set of hard constraints. 
You must assess both the numeric durations and logical feasibility of the plan. Be strict and precise.
(Optional) \{Verification steps\}\\

Given query, Please evaluate whether the plan satisfies all the constraints.

\textbf{Query}:\\
\{query\}\\

\textbf{Input plan}:\\
\{solution\} \\

\textbf{Constraints}: \\
\{constraints\} \\

Format your response strictly as follows - no extra text, comments, or explanations:\\
Score: [integer number, reward score between 0 and 100]\\
Violated Constraints: [string, list any constraints that the plan violates or any errors in the plan, and provide the reason why each constraint was violated]
\end{tcolorbox}
\clearpage
\begin{tcolorbox}[title=Actor Prompt,
  colback=gray!0,   
  colframe=black!65, 
  fonttitle=\bfseries,
  width=1.1\textwidth,
  valign=center, 
]
You are an expert at planning problems. \{dataset specific description\}.\\
Along with the task, you are also given 5 example tasks and solutions. You can learn from the examples to generate your own plan on the given task.\\
Follow the format of the example solution and the required output format precisely, starting with \{dataset specific description\}. \\
Only provide the travel plan, no other text. \\

\textbf{Query}:\\
\{query\}\\

\textbf{Constraints}: \\
Your solution must strictly satisfy the following constraints.\\
\{constraints\} \\
If any constraint is violated, the solution is considered incorrect \\

\textbf{Previous Errors}:\\
The following notes lists why previous attempts failed: \\
\{errors\} \\
Do not repeat the same errors. \\

Based on all the information above, generate your final output.
\end{tcolorbox}

\vspace{3em}

\section{Examples of Different Agents in EvoMem}
\label{app:examples}

\begin{table}[ht]
\centering
\begin{tblr}{
  colspec = {X}, 
  hline{1,Z} = {0.08em},           
  hline{2}   = {0.05em},           
  row{1} = {font=\bfseries, halign=c}, 
  rowsep = 1.5pt,
}
Constraint Extractor (Trip Planning) \\

\textbf{Problem:}\\
You plan to visit 3 European cities for 14 days in total.\\
You only take direct flights to commute between cities.\\
You would like to visit Florence for 6 days.\\
You want to meet a friend in Florence between day 9 and day 14.\\
You would like to visit Barcelona for 5 days.\\
You would like to visit Helsinki for 5 days.\\
Here are the cities that have direct flights: Barcelona and Florence; Helsinki and Barcelona.\\

\textbf{Generated Constraints:}\\
1.\; Total trip duration: 14 days\\
2.\; Visit durations per city:\\
\quad-- Florence: 6 days (Day 9–14)\\
\quad-- Barcelona: 5 days\\
\quad-- Helsinki: 5 days\\
3.\; Fixed-date events: meet friend in Florence between Day 9–14\\
4.\; Flight constraints:\\
\quad-- Direct flights only\\
\quad-- Direct flight connections: Barcelona–Florence, Helsinki–Barcelona\\
\end{tblr}
\vspace{6pt}
\caption{Example input and extracted constraints for the constraint extractor in trip planning.}
\end{table}

\begin{table}[ht]
\centering
\begin{tblr}{
  colspec = {X}, 
  hline{1,Z} = {0.08em},        
  hline{2}   = {0.05em},        
  row{1}    = {font=\bfseries, halign=c}, 
  rowsep=1pt,
}
Constraint Extractor (Calendar Scheduling) \\

\textbf{Problem:}\\
You need to schedule a meeting for Michelle, Steven, and Jerry for one hour between the work hours of 9:00 to 17:00 on Monday.\\
Here are the existing schedules for everyone during the day:\\
~- Michelle has meetings on Monday during 11:00–12:00;\\
~- Steven has blocked their calendar on Monday during 9:00–9:30, 11:30–12:00, 13:30–14:00, 15:30–16:00;\\
~- Jerry has blocked their calendar on Monday during 9:00–9:30, 10:00–11:00, 11:30–12:30, 13:00–14:30, 15:30–16:00, 16:30–17:00.\\

\textbf{Generated Constraints:}\\
1.~ Meeting Participants: Michelle, Steven, Jerry\\
2.~ Meeting Duration: 1 hour (60 minutes)\\
3.~ Meeting Day: Monday\\
4.~ Allowed Meeting Hours: 09:00–17:00\\
5.~ Blocked Schedules:\\
\quad-- Michelle: 11:00–12:00\\
\quad-- Steven: 09:00–09:30, 11:30–12:00, 13:30–14:00, 15:30–16:00\\
\quad-- Jerry: 09:00–09:30, 10:00–11:00, 11:30–12:30, 13:00–14:30, 15:30–16:00, 16:30–17:00\\
6.~ Meeting Preferences: None\\
\end{tblr}
\vspace{6pt}
\caption{Example input and extracted constraints for the constraint extractor in calendar scheduling.}
\end{table}

\begin{table}[ht]
\centering
\begin{tblr}{
  colspec = {X}, 
  hline{1,Z} = {0.08em}, 
  hline{2}   = {0.05em}, 
  row{1}    = {font=\bfseries, halign=c},
  rowsep=1pt,
}
Constraint Extractor (Meeting Planning) \\

\textbf{Problem:}\\
You are visiting San Francisco for the day and want to meet as many friends as possible. \\
Solve the problem by considering various schedules and picking the best one to optimize your goals.\\
Travel distances (in minutes):\\
~- Marina District → Mission District: 20\\
~- Mission District → Marina District: 19\\
You arrive at Marina District at 9:00AM. \\
Stephanie will be at Mission District from 10:30AM to 1:30PM. \\
You'd like to meet Stephanie for a minimum of 120 minutes. \\

\textbf{Generated Constraints:}\\
1.~ Arrival Time Constraint:\\
\quad-- You arrive at Marina District at 9:00 AM\\
2.~ Travel Time Constraints:\\
\quad-- Marina District → Mission District: 20 minutes\\
\quad-- Mission District → Marina District: 19 minutes\\
3.~ Availability Constraint:\\
\quad-- Stephanie is at Mission District from 10:30 AM–1:30 PM\\
4.~ Meeting Duration Constraint:\\
\quad-- You want to meet Stephanie for at least 120 minutes\\
5.~ Optimization Goal:\\
\quad-- Maximize the number of friends met during the day by considering various schedules.\\
\end{tblr}
\vspace{6pt}
\caption{Example input and extracted constraints for the constraint extractor in meeting planning.}
\end{table}

\clearpage

\begin{table}[ht]
\centering
\begin{tblr}{
  colspec = {X},
  hline{1,Z} = {0.08em},        
  hline{2}   = {0.05em},        
  row{1}     = {font=\bfseries, halign=c},
  rowsep= 0.4pt,
}
Verifier (Trip Planning) \\

\textbf{Constraints:}\\
1.\; Total trip duration: 12 days\\
2.\; Visit durations per city:\\
\quad-- Tallinn: 2 days (Day 11–12)\\
\quad-- Seville: 7 days\\
\quad-- Munich: 5 days\\
3.\; Timing constraints: meet friend in Tallinn between Day 11–12\\
4.\; Flight constraints:\\
\quad-- Direct flights only\\
\quad-- Direct flight connections: Seville–Munich, Munich–Tallinn\\[2pt]

\textbf{Solution:}\\
Here is the trip plan for visiting the 3 European cities for 12 days:\\
\quad\textbf{Day 1–7:} Arrive in Seville and visit for 7 days.\\
\quad\textbf{Day 7:} Fly from Seville to Munich.\\
\quad\textbf{Day 7–10:} Visit Munich for 5 days.\\
\quad\textbf{Day 10:} Fly from Munich to Tallinn.\\
\quad\textbf{Day 10–12:} Visit Tallinn for 2 days.\\

\textbf{Score:} 90\\
\textbf{Errors:} Munich visit duration is incorrect: the plan states 5 days but the itinerary covers 4 days (Day 7–10).\\
\end{tblr}
\vspace{6pt}
\caption{Verifier example for trip planning.}
\end{table}
\begin{table}[ht]
\centering
\begin{tblr}{
  colspec = {X},
  hline{1,Z} = {0.08em},        
  hline{2}   = {0.05em},        
  row{1}     = {font=\bfseries, halign=c},
  rowsep=0.4pt,
}
Verifier (Calendar Scheduling) \\

\textbf{Constraints:}\\
1.\; Meeting Participants: Arthur, Michael, and Samantha\\
2.\; Meeting Duration: 1 hour (60 minutes)\\
3.\; Meeting Day: Monday\\
4.\; Allowed Meeting Hours: Between 09:00 and 17:00\\
5.\; Blocked Schedules:\\
\quad-- Arthur: 09:00–09:30, 10:30–12:00, 16:00–17:00\\
\quad-- Michael: 13:00–13:30, 14:00–14:30\\
\quad-- Samantha: 10:30–11:00, 12:00–15:00, 15:30–17:00\\
6.\; Meeting Preferences: None\\
7.\; Feasibility: True (explicitly stated “there exists a solution”)\\

\textbf{Solution:}\\
Proposed meeting time: Monday, 12:00–13:00\\

\textbf{Score:} 0\\
\textbf{Errors:} Samantha is busy from 12:00 to 15:00, overlapping with the proposed meeting time of 12:00–13:00.\\
\end{tblr}
\vspace{6pt}
\caption{Verifier example for calendar scheduling}
\end{table}

\clearpage

\begin{table}[ht]
\centering
\begin{tblr}{
  colspec = {X},
  hline{1,Z} = {0.08em},        
  hline{2}   = {0.05em},        
  row{1}     = {font=\bfseries, halign=c},
  rowsep=2pt,
}
Verifier (Meeting Planning) \\

\textbf{Constraints:}\\
1.\; Arrival Time Constraint:\\
\quad-- You arrive at Chinatown at 9:00 AM\\
2.\; Travel Time Constraints:\\
\quad-- Chinatown → Russian Hill: 7 minutes\\
\quad-- Russian Hill → Chinatown: 9 minutes\\
3.\; Availability Constraint:\\
\quad-- Ronald is at Russian Hill from 3:15 PM to 9:30 PM\\
4.\; Meeting Duration Constraint:\\
\quad-- You want to meet Ronald for at least 105 minutes\\
5.\; Optimization Goal:\\
\quad-- Maximize the number of friends met during the day by considering various schedules\\[2pt]

\textbf{Solution:}\\
1.\; Arrival Time Constraint:\\
\quad-- You arrive at Chinatown at 9:00 AM\\
2.\; Travel Time Constraints:\\
\quad-- Chinatown → Russian Hill: 7 minutes\\
\quad-- Russian Hill → Chinatown: 9 minutes\\
3.\; Availability Constraint:\\
\quad-- Ronald is at Russian Hill from 3:15 PM to 9:30 PM\\
4.\; Meeting Duration Constraint:\\
\quad-- You want to meet Ronald for at least 105 minutes\\
5.\; Optimization Goal:\\
\quad-- Maximize the number of friends met during the day by considering various schedules\\[2pt]

\textbf{Score:} 90\\
\textbf{Errors:} Meeting Duration Constraint: The meeting with Ronald only lasts 45 minutes (from 3:15 PM to 4:00 PM), which is less than the required minimum duration of 105 minutes.\\
\end{tblr}
\vspace{6pt}
\caption{Verifier example for meeting planning: constraints, proposed solution, score, and detected errors.}
\end{table}

\newpage

\end{document}